\documentstyle[12pt]{article}
\input epsf

\input epsf

\begin{document}

\renewcommand{\thesection}{\arabic{section}.} 
\renewcommand{\theequation}{\thesection \arabic{equation}}
\newcommand{\scs}{\setcounter{equation}{0} \setcounter{section}}
\def\req#1{(\ref{#1})}
\newcommand{\be}{\begin{equation}} \newcommand{\ee}{\end{equation}} 
\newcommand{\ba}{\begin{eqnarray}} \newcommand{\ea}{\end{eqnarray}} 
\newcommand{\la}{\label} \newcommand{\nb}{\normalsize\bf} 
\newcommand{\lb}{\large\bf} \newcommand{\vol}{\hbox{Vol}}
\newcommand{\bb} {\bibitem} \newcommand{\np} {{\it Nucl. Phys. }} 
\newcommand{\pl} {{\it Phys. Lett. }} 
\newcommand{\pr} {{\it Phys. Rev. }} \newcommand{\mpl} {{\it Mod. Phys. Lett. }}
\newcommand{\sg}{{\sqrt g}} \newcommand{\sqhat}{{\sqrt{\hat g}}}
\newcommand{\sqphi}{{\sqrt{\hat g}} e^\phi} 
\newcommand{\sqalpha}{{\sqrt{\hat g}}e^{\alpha\phi}}
\newcommand{\tp}{\cos px\ e^{(p-{\sqrt2})\phi}} \newcommand{\stwo}{{\sqrt2}}
\newcommand{\tr}{\hbox{tr}}

\begin{titlepage}
\renewcommand{\thefootnote}{\fnsymbol{footnote}}

\hfill BUTP-97/37

\hfill hep-th/yymmxxx

\vspace{.4truein}
\begin{center}
 {\LARGE Wrapped Branes and Confined Momentum}
 \end{center}
\vspace{.7truein}

 \begin{center}

 Christof Schmidhuber\footnote{christof@butp.unibe.ch; work supported by Schweizerischer Nationalfonds.}

 \vskip5mm

 {\it Institute of Theoretical Physics, Sidlerstr. 5, 3012 Bern, Switzerland}

 \end{center}

\vspace{1truein}
\begin{abstract}

We present string-like soliton solutions of three-dimensional gravity, coupled to a compact
scalar field $x^{11}$ and Kaluza-Klein reduced on a circle. These solitons carry fractional
magnetic flux with respect to the Kaluza-Klein gauge field. 
Summing over such ``Kaluza-Klein flux tubes'' is shown to 
imply summing over a subclass of three-dimensional topologies (Seifert manifolds).
It is also shown to imply an area law for the Wilson loop of the Kaluza-Klein gauge field; the
confined charge is nothing but Kaluza-Klein momentum. 
Applied to the membrane of M-theory, this is interpreted as ``dynamical wrapping'' of the M-brane
around its eleventh embedding dimension $x^{11}$.

\end{abstract}
 \renewcommand{\thefootnote}{\arabic{footnote}}
 \setcounter{footnote}{0}
\end{titlepage}

{}\subsection*{1. Introduction}\scs{1}

String theory is no longer a theory of only strings (one-branes), but
also contains various other dynamical $p$-branes.
In particular, the type IIA theory in ten dimensions
contains a two-brane.
In the original formulation of string theory, where the one-branes
are regarded as fundamental objects, the two-brane appears as a Dirichlet brane \cite{polchinski}
- the stringy version of a soliton.
In a dual formulation (Matrix theory \cite{susskind}), that regards the zero-branes
of type IIA string theory as fundamental objects,
the two-brane appears as a bound state.

No matter how the theory is formulated, one would like to better understand
the dynamics of the two-brane. Here we would like to 
present some observations which seem to suggest that
the standard theory of dynamical one-branes can be embedded in a theory of dynamical two-branes.
This involves applying an old concept of field theory - confinement - to
gravity, and it might explain why branes wrap in M-theory.

Dynamical one-branes of given topology
are very successfully described by renormalizable theories of two-dimensional
quantum gravity coupled to matter \cite{polyakov2d}. Moreover, orientable one-brane-topologies are easily
classified by their genus, and there is a topological expansion parameter
- the string coupling constant - that can be used to pick out the simplest topology
- the two-sphere - to start with. Then the other topologies can be added as perturbations.
But there are at least two outstanding
problems in extending this approach to the three-dimensional two-branes:
\begin{enumerate}
\item
The three-dimensional Hilbert-Einstein action should appear as a counterterm
in the world-brane action. But
$3d$ gravity coupled to matter is not a renormalizable theory.
\item
There is no topological expansion
parameter that can be used to restrict attention to a single simple
world-brane topology, such as $S^3$ or $S^2\times S^1$.
\end{enumerate}
One might try to avoid the first problem by fine-tuning the Hilbert-Einstein action away,
or by hoping that it is excluded by an unknown nonrenormalization theorem. But
even if this does solve the first problem, it does not solve the second one.
What we would like to suggest instead is that
summing over three-dimensional world-brane topologies has the potential of reducing
the world-brane theory to a renormalizable two-dimensional one.

In the absence of a classification of general three-dimensional topologies we will 
restrict attention to a certain - still very rich - subset of topologies
that includes both $S^3$ and $S^2\times S^1$: Seifert manifolds \cite{seifert}. 
Those are the manifolds that admit a foliation by circles, and they can be completely
classified. We will argue that, within this subset, summing over topologies has a
dramatic effect: it leads to linear confinement of Kaluza-Klein momentum in the circular direction.
This is a lower-dimensional version of previous suggestions by Gross \cite{gross} and Witten \cite{wittencom}.
We will then offer a speculative interpretation of confinement of Kaluza-Klein
momentum in terms of ``wrapping'' of two-branes.

In section 2, properties of the two-brane of type IIA string theory in 10 dimensions
are recalled. The auxiliary world-brane metric is introduced and an induced three-dimensional
Hilbert-Einstein term is included in the action.
In sections 3 and 4, this theory is considered on a two-brane of topology $S^2\times S^1$.
It is shown that the classical equations of motion 
have finite-action solutions that describe string solitons which are wrapped around the $S^1$.
We call them Kaluza-Klein flux tubes, because they carry fractional magnetic flux
with respect to the Kaluza-Klein gauge field that arises from the world-brane
metric upon compactification along the $S^1$. These flux tubes involve vortex lines
of the eleventh dimension of M-theory; we argue that such vortex lines are allowed
on the membrane, even though vortices are forbidden on a string world-sheet.
We then discuss two effects of these flux tubes:

\begin{itemize}\item
In section 5, it is shown that the Kaluza-Klein flux tubes change the topology of the
two-brane by performing what is called ``Dehn surgery'' on it.
A single flux tubes changes the topology from $S^2\times S^1$ to a lens space,
and by summing over an arbitrary number of flux tubes one sums over all Seifert manifolds.
\item
In section 6, the Wilson loop for the Kaluza-Klein gauge field is computed.
It is seen to obey an area law. This implies linear confinement of the associated
electric charge, which is nothing but Kaluza-Klein momentum. 
\end{itemize}
In the latter computation
we take the semiclassical limit, where the three-dimensional Newton constant goes to
zero. In this limit we define the path integral simply as a sum over soliton configurations.
Because of unboundedness problems one should really study the supersymmetric
version of the theory where the Kaluza-Klein flux tubes are BPS-saturated;
this is under investigation.

Section 7 offers a speculative interpretation of confinement of Kaluza-Klein momentum:
it is argued to imply that two-branes are ``permanently compactified'' (or ``wrapped'') 
to one-branes,
similarly as quarks are permanently confined in QCD. Moreover, it is argued
that neutral bound states of Klauza-Klein modes (``baryons'') cannot exist
and that, as a consequence, these one-branes are standard strings.
Comments on the nonabelian generalization of these proposals
conclude the paper.

Some of these ideas have appeared in a previous publication by the author at the example of
a simplified toy model \cite{ichtop}.

{}\subsection*{2. Dynamical membranes}\scs{2}

Let us begin by recalling some of the properties of the solitonic two-brane of
type IIA string theory in ten dimensions. These properties can be computed directly using
techniques of open string theory \cite{polchinski,leigh,clny,ichD}. 
Among the results that one finds are the following:

The world-brane fields that live on the two-brane
are the space-time coordinates $x^\mu$ with $\mu\in\{1,...,10\}$, as well as an
eleventh embedding dimension $x^{11}$, which is the dual of
the world-brane U(1) gauge field \cite{duff, townsend, ichD}. 
An important property of $x_{11}$ is that it is compact. The
compactification radius $R$ 
is related to the string coupling constant $\kappa$ \cite{witten11}:
$$ x_{11}\ \equiv\ x_{11}+2\pi R\ ,\ \ \ R\sim\kappa^{2\over3}\ .$$
The world-brane action of the Dirichlet two-brane comes out to be
the eleven-dimensional supermembrane action \cite{bst}
whose bosonic part is 
\ba
S\ =\ T \int d^3\sigma\ {\sqrt{\det\ G_{ij}}}\ +\ \hbox{Wess-Zumino term}\ .
\la{bettina}\ea
Here $i,j\in \{1,2,3\}$ and the world-brane and space-time signatures are
taken to be Euclidean.
$G_{ij}$ is the 
induced world-brane metric $\partial_i\vec x\partial_j\vec x$ 
with $\vec x\equiv(x^\mu,x^{11})$.
An interesting aspect of the two-brane action is that it has eleven-dimensional
general covariance. So although type IIA string theory was defined purely in ten
dimensions, its solitonic two-brane really thinks she lives in eleven dimensions and she is
the supermembrane.

The world-brane action (\ref{bettina}) is in Nambu-Goto form.
Because of their complicated forms,
Nambu-Goto actions are not very useful when one tries to quantize branes.
So we rewrite (\ref{bettina}) with the help of a $3d$ metric $h_{ij}$ \cite{tucker} as
\ba
S\ =\ {T\over2} \int d^3\sigma\ {\sqrt{h}}\{h^{ij}\partial_i\vec x\partial_j\vec x \ -\ 1 \}\ ,
\la{diana}\ea
Classically, the saddle point value of the integral over $h$ at 
$h_{\alpha\beta} = \partial_\alpha x^\mu\partial_\beta x_\mu$ reproduces 
the Nambu-Goto action.

Quantum mechanically, we must add to the Lagrangean
counterterms that are induced in the process of renormalizing this
theory of three-dimensional gravity coupled to matter fields $x^A$.
Among them is - at least for the bosonic membrane - the Hilbert-Einstein term
\ba -{1\over 2e^2}\int {\sqrt{h}}{\cal R}^{(3)}\ .\la{beate}\ea
where ${\cal R}^{(2)}$ is the two-curvature and $e$ is some coupling constant. But this
leads to a nonrenormalizable theory of three-dimensional 
gravity coupled to matter that really does not seem to make any sense as a quantum theory.
As mentioned in the introduction, it is not clear how to make sense of the theory
even without the Hilbert-Einstein action, i.e. in the limit $e\rightarrow\infty$.

So here we will consider the theory instead in the semiclassical limit $e\rightarrow0$,
where we define the path integral as a sum over classical solutions of the Einstein equations.
In section 4 we will find a set of nontrivial classical solutions.
Specifically, we will consider membranes of topology
$$R^2\ \times S^1\ .$$
Let us parametrize the $S^1$ by the coordinate $z\in[0,2\pi[$, and the $R^2$ 
by coordinates $\sigma_1,\sigma_2$.
If the membrane wraps $n$ times around $x^{11}$, this means that $x^{11}=nRz$.
In the solutions of section 4, all the other coordinates $x^\mu$ as well as the
world-brane metric $h_{ij}$ depend 
only on $\sigma_1,\sigma_2$; $x^{11}$ may also have a piece that depends on
$\sigma_1,\sigma_2$:
\ba
h_{ij}&=&h_{ij}(\sigma_1,\sigma_2)\la{miriam}\\
x^\mu&=& x^\mu(\sigma_1,\sigma_2)\\  
x_{11}&=& nRz\ +\ \tilde x_{11}(\sigma_1,\sigma_2)\ .
\la{clara}\ea
In this case it is useful to perform a standard Kaluza-Klein reduction of the three-dimensional
metric to a two-dimensional metric $h_{\alpha\beta}$
on $R^2$, a two-dimensional Kaluza-Klein gauge field $A_\alpha$
and a scalar $L$, which measures the size of the circular world-brane direction.
They are defined by the line element
\ba
(ds)^2\ =\ h^{(2)}_{\alpha\beta}\ d\sigma^\alpha d\sigma^\beta \ +\ 
L^2(dz+A_\alpha d\sigma^\alpha)^2\ .
\la{elise}\ea
The three-dimensional Ricci scalar becomes:
$$
 {\cal R}^{(3)} \ \rightarrow\ {\cal R}^{(2)} - {1\over2}L^2F_{\alpha\beta}^2\ ,
$$
where $F_{\alpha\beta}$ is the field strength of $A_\alpha$.
Then the action (\ref{diana} plus \ref{beate}) becomes:
\ba
 S\ \ \sim\ \ \int d^2\sigma\ L\sqrt{ h ^{(2)}}&\times& \{\ {T\over2}h^{\alpha\beta}\partial_\alpha x^\mu 
  \partial_\beta x_\mu \ -\ {1\over 2e^2} {\cal R}^{(2)}\la{erika}\\ 
&&+ {T\over2}h^{\alpha\beta}(\partial_\alpha x_{11}-nRA_\alpha )
  (\partial_\beta x_{11}-nRA_\beta)  + {1\over4e^2}L^2F_{\alpha\beta}^2\\
&&+ {T\over2}({n^2R^2\over L^2}\ -\ 1)\ \}\la{flavia}\ .
\la{gabi}\ea
Up to the overall factor $L$, the first line resembles a standard Nambu-Goto string embedded
in ten dimensions. 
M-theory adds to this string the gauge field and $x^{11}$ appearing in the second line (plus a potential 
term). 
What M-theory adds seems to be trivial at first sight - the
gauge field seems to simply ``eat'' $x^{11}$. However, $x_{11}$ is compact and therefore
there might be vortices in the system.

{}\subsection*{3. Are vortex lines of $x^{11}$ allowed?}\scs{3}

Let us now parametrize the two-brane of topology $R^2\times S^1$
by the circular coordinate $z$ and by polar coordinates $(r,\phi)$
in $R^2$. By a Kaluza-Klein flux tube centered at $r=0$,
we mean a vortex configuration that obeys in addition to (\ref{miriam}-\ref{clara}):
\ba
x^{11}&=&mR\phi\ +\ nRz \la{rebecca} \\
 A_\phi &\rightarrow& {m\over nr}\ \ \ \hbox{for}\ \ \ r\rightarrow\infty\la{rosi}\\
 A_\phi &\rightarrow& 0\ \ \ \ \ \hbox{for}\ \ \ r\rightarrow0\ ,\la{ruth}
\ea
where the Kaluza-Klein gauge field has been assumed to only have an angular component $A_\phi$.
Such a vortex obviously encloses fractional magnetic flux 
$$\int F\ =\ 2\pi{m\over n}\ .$$

Before finding solutions of the equations of motion, let us ask whether we are really
permitted to include such vortex configurations in the path integral.
There are two reasons why one may have doubts. First one may ask,
isn't there a Dirac quantization rule that forces the magnetic flux to be integer?
We will return to this in section 5, where we will show that in the case at hand the
Dirac condition indeed only requires the flux to be $2\pi$ times a {\it rational} number.

Second, one may think that vortex lines of $x^{11}$ are forbidden on the membrane
for the same reason that vortices of compact coordinates are forbidden on a string
world-sheet. Suppose there is a compact coordinate $x^1$ with radius $R$ in string theory.
Vortices
$$x^1\ =\ mR\phi$$
on the string world-sheet are forbidden, because they create holes in the world-sheet:
an infinitesimal circle $\zeta$ drawn on the world-sheet around the vortex center is mapped
in target space onto a line of finite length $2\pi mR$, so the world-sheet acquires
a boundary. One consequence of this is that gauge invariance of the Neveu-Schwarz
two-form of string theory is lost \cite{wittenbound}: the vortex is a source
of violation of the gauge invariance
$$B_{1\mu}\ \rightarrow\ B_{1\mu}+\partial_\mu \Lambda\ ,$$
where $\Lambda$ is an arbitrary space-time scalar field, since
$$\delta\int B\ 
\equiv\ \delta\int\epsilon^{\alpha\beta}\partial_\alpha x^\mu\partial_\beta x^\nu B_{\mu\nu}
\ =\ \oint_\zeta\Lambda d x_1\ =\ 2\pi mR\ \Lambda(r=0)\ .$$
(By $dx_1$ we mean $\partial_\alpha x_1 d\sigma^\alpha$.)
Another way of putting things is, vortices are vertex operators for string winding modes.
But in string theory we only allow vertex operators and target space fields
for momentum modes and not for winding modes, for the reason just mentioned.\footnote{I
thank Joe Polchinski for making this point.}

Similarly, one may wonder whether vortex lines on the membrane spoil
gauge invariance of the three-form gauge field of M-theory,
$$C_{11\mu\nu}\ \rightarrow\ C_{11\mu\nu}+\partial_{[\mu}\Lambda_{\nu]}\ .$$
One can easliy check that this is not the case. If $\zeta\times S^1$ is a thin torus
enclosing the vortex line, then
$$\delta\int C\ =\ 2\pi nR\oint_{\zeta\times S^1} \Lambda_\nu dx^\nu\wedge dx^{11}\ \rightarrow\ 0\ $$
as $\zeta$ shrinks to a point. The intuitive reason is that
vortex lines (\ref{rebecca}) do {\it not} create boundaries of the membrane. 
In fact, they can locally be absorbed in a large diffeomorphism such as
$$z\ \rightarrow\ z+m\phi\ $$
for $n=1$ (the case $n\neq1$ will be discussed below).

Since vortices can locally be absorbed in large diffeomorphisms, one may think that they should still 
be excluded, as they are ``pure gauge''. But the point is
that they cannot be absorbed globally for a general world-brane topology,
e.g. for a lens space. This will become clear below.
So there does not seem to be an objection of principle against vortices of $x_{11}$.
We will therefore now discuss solutions of the equations of motion that involve them.

{}\subsection*{4. Kaluza-Klein flux tubes}\scs{4}

To find solutions with boundary conditions (\ref{rebecca}-\ref{ruth}), we
make the following ansatz for the $3d$ metric:
$$ds^2\ =\ dr^2\ +\ \rho(r)^2d\phi^2\ +\ L(r)^2(dz^2+\rho(r) A(r)d\phi)^2\ .$$
In comparing with (\ref{elise}), note that $A$ is here defined to be accompanied by $\rho$.
The field strength of the Kaluza-Klein gauge field $A$ is then
$$ F(r)\ =\ {(\rho A)'\over \rho} .$$
We also assume that the membrane is ``stretched'', i.e. the $r-\phi-$plane
is identified with the $x_1-x_2-$plane:
\ba
x_1&=& f(r)\cos\phi\\ x_2&=& f(r)\sin\phi\ .\ea
Because of the presence of the vortex, the function
$f(r)$ cannot be set equal to $r$ but must be determined from the equations of motion. 
It is straightforward to compute the spin connection, the curvature and from it the 
equations of motion of the action (\ref{diana} plus \ref{beate}). For the 
equations of motion we find:
\ba
f''\ +\ {(\rho L)'\over\rho L }f'\ -\ {1\over\rho^2}f&=&0 \la{zero}\\
\ -{1\over e^2T}\{{\rho''\over\rho}+{L''\over L}+{L^2F^2\over2}\}&=& (f')^2\ -\ 1\la{one}\\
\ -{1\over e^2T}\{{\rho''\over\rho}+{\rho'\over\rho}{L'\over L}+{L^2F^2\over2}\}&=&R^2\ 
   ({m\over\rho}-nA)^2\ +\ {f^2\over\rho^2}\ -\ 1\la{two}\\
\ -{1\over e^2T}\{{L''\over L}+{\rho'\over\rho}{L'\over L}-{L^2F^2\over2}\}&=&R^2\ {n^2\over L^2}\ -\ 1
\la{three}\\
\ -{1\over e^2T}\{3L'F+LF'\}&=&R^2\ {2n\over L}\ ({m\over \rho}-nA) \ \la{four}.
\ea
Only four of these five equations are independent. In fact,
a linear combination of them,
(\ref{one}) minus (\ref{two}) minus (\ref{three}), is a constraint
that is first-order in derivatives. Its derivative is implied by
the other equations. 
Up to this constraint, the equations of motion are those of action (\ref{erika}-\ref{flavia}),
reduced to one dimension:
\ba S&=&{4\pi^2\over e^2}\int dr\  \{{\rho L^3\over4}F^2\ -\rho'L'\}\ +\ {4\pi^2\over e^2}[L\rho']^\infty_0
\la{uli1}\\
 &+& 2\pi^2 T\int dr\  \rho L\ \{\ {n^2\over L^2}R^2\
+\ ({m\over\rho}-nA)^2R^2\ +\ (f')^2\ +\ {f^2\over\rho^2}\ -\ 1\ \}\ .\la{uli2}\ea
In the first line we have used the fact that ${\sqrt {g^{(3)}}}R^{(2)}=-2L\rho''$ and 
kept the boundary term.

For $m=0$, the equations of motion have the trivial solution
\ba f(r)\ =\ \rho(r)\ =\ r\ ,\ \ L\ =\ nR\ ,\ \ A\ =\ 0\ .\la{sandy}\ea
For $m\neq0$ we have only been able to find analytic solutions near the origin and at infinity,
but Mathematica has been able to find well-behaved vortex solutions everywhere; an example
is shown in Fig.1. There, $(\rho A),L,\rho$ and $f$ are plotted as functions of $r$ for $e=R=1$ 
and $n=5,m=2$.\footnote{For anyone 
who would like to reproduce the figure: the ``initial conditions'' were 
roughly determined from the asymptotic solution near $r=0$ (given below) and then 
fine-tuned to satisfy the boundary conditions at infinity (given below). They were fine-tuned
at $r=0.000123$: 
$\rho A=0, (\rho A)'=0.0002307{2\over 5z^2}, L=15z, 
L'=-6500 z, r=0.00074, f=0.00143 f_0, f'=3.87 f_0$ with $z=0.626055, f_0=72.996.$
By fine-tuning, one can simultaneously match the boundary conditions at zero and infinity with arbitrary 
precision.}

\begin{picture}(500,300) 
\put(10,260){$\rho(r)A(r)$}  
\put(250,260){$\rho(r)$}  
\put(20,110){$L(r)$}
\put(250,110){$f(r)$}  
\put(180,165){$r$}  
\put(180,40){$r$}  
\put(425,165){$r$}  
\put(425,20){$r$}  
\end{picture}

 {}\epsffile[-5 5 0 0]{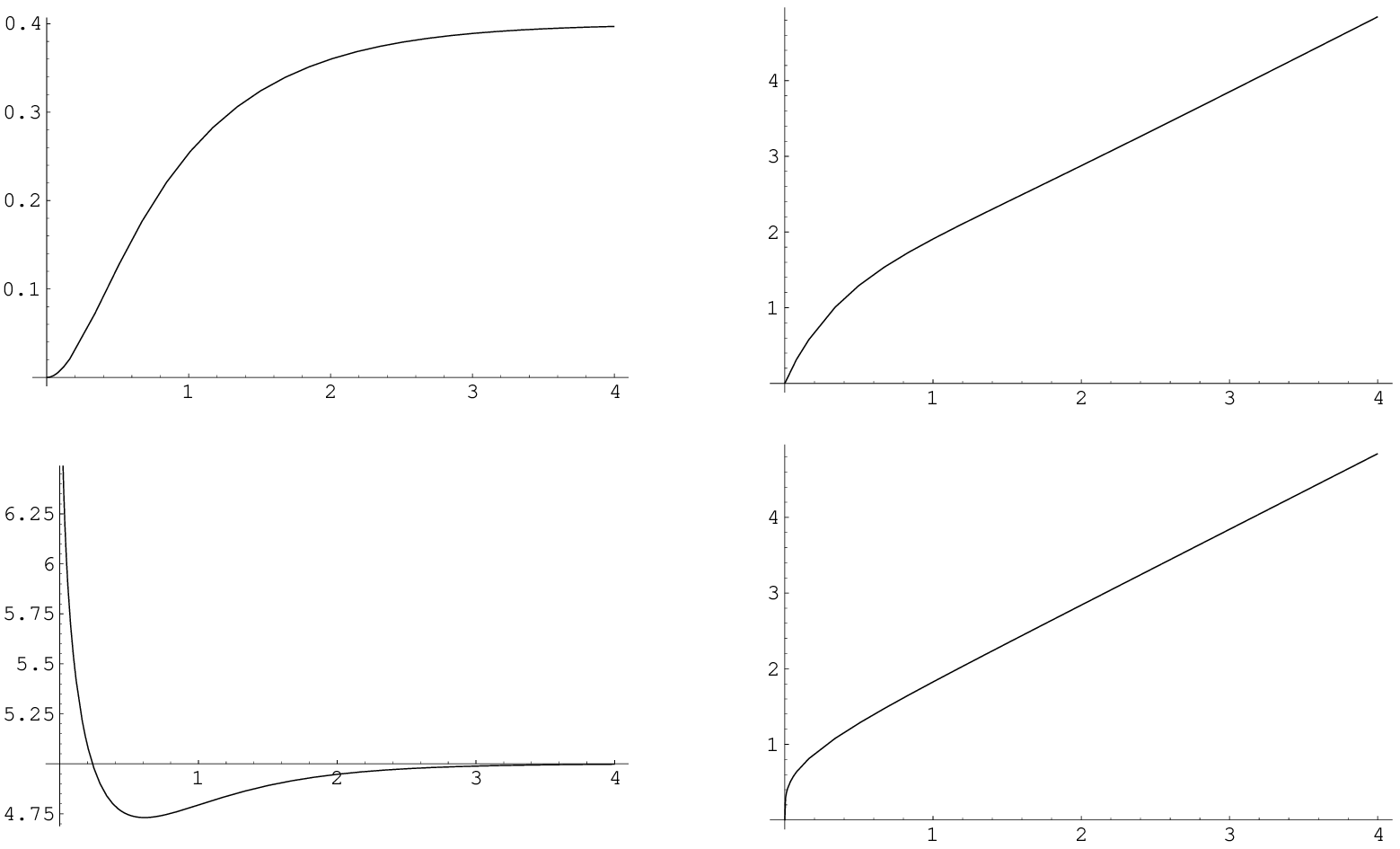}

\begin{center} {\small Fig. 1: Example of a Kaluza-Klein flux tube with n=5, m=2.  }\end{center}
\vskip.5cm

These solutions have the following asymptotic behavior. At $r\rightarrow\infty$:
\ba L&=&nR\ +\ ...\\
\rho&=&r + r_0\ +\ ...\\
A&=&{m\over n(r+r_0)}\ +\ ...\\
f&=&r + r_0\ +\ ...,\  
\ea
where the dots represent terms of order $e^{-{\sqrt{2T}}er}$.
Here $r_0$ is a constant that could be absorbed in a shift $r\rightarrow r-r_0$; 
then the vortex center 
would be at $r=r_0$.  Thus the vortex is classically invisible far away from its core, in the sense that
not only the magnetic field vanishes but also $L, f$  and $\rho$ assume the same values as the solution
(\ref{sandy}) with $m=0$.\footnote{But there is a quantum mechnaical Aharonov-Bohm effect; see section 5.}
Note, in particular, that there is no deficit angle at infinity.

To achieve this asymptotic behavior at infinity, it suffices to impose three boundary conditions:
$$  \rho A\rightarrow{m\over n}
\ \ ,\ \ \ \ \ L'\rightarrow0\ \ , \ \ \ \ \ f'\rightarrow1\ \ \ \ \ \ \ \ \ \hbox{at}\ r\rightarrow\infty\ .$$
The rest follows from the equations of motion.

Near the origin, the solutions behave as follows:
\ba L&=& nR{\sqrt T}e\lambda\ |\log er|^{1\over2}+...\la{maja1} \\
   \rho&=& |m|R{\sqrt T}\ er|\log er|^{1\over2}+... \\
   \rho A&=& {2\over3}{m\over n}{r^2\over\lambda^2} +...\\
   f&=& f_0|\log er|^{-{1\over4}}\exp\{-{2\over |m|eR{\sqrt T}}|\log er|^{1\over2}\}+...\la{maja4} \ea
with free parameters $\lambda$ and $f_0$.
This behavior follows from the equations of motion if three more boundary conditions are imposed:
$$ \ A(r)\rightarrow0\ \ ,
\ \ \ \ \ \rho(r)\rightarrow0\ \ ,\ \ \ \ \ f(r)\rightarrow0\ \ \ \ \ \ \ \ \ \hbox{at}\ r\rightarrow0\ .$$
The second and third conditions insure that the vortex does not create a
hole in the membrane, neither in the internal nor in the embedding space.\footnote{To see that
(\ref{maja1}-\ref{maja4}) solve the equations of motion near $r=0$, note that
these expressions obey
$L'=-{1\over2}n|m|\lambda e^2R^2T{1\over\rho}; \rho'=|{m\over n}|{1\over\lambda}L;f'={f\over\rho};
F=\pm{4\over3\lambda},$ neglecting terms that are suppressed by negative powers of $|\log er|$
and by exp\{$-{\sqrt{\log er}}$\}. Under the same approximations, these first order equations
imply the equations of motion (\ref{zero}-\ref{four}).}

For these solutions, the ``brane thickness '' $L$ diverges very slowly at the origin.
There is also a curvature singularity at the origin, in the sense that the
deficit angle $-2\pi\rho'|^{r=\infty}_{r=0}$ also diverges very slowly there:
$$2\pi\ \rho'(r)|^{r=\infty}_{r=0}\ \rightarrow\ 2\pi |m|eR{\sqrt T}
\ |\log er|^{1\over2}\ \ \ \ \ \ \hbox{as}\ \ r\rightarrow0\ .$$
This divergence causes no problems, since the flux tube action nevertheless turns out to be finite:
it is easy to check that on solutions of the equations of motion action (\ref{uli1}-\ref{uli2}) becomes
a boundary term plus a volume term:
\ba S\ =\ -{4\pi^2\over e^2} [\rho L']^{r=\infty}_{r=0} \ +\ 4\pi^2T\int dr\ \rho L\ .\la{marina}\ea
The second term is, after shifting $r$, the action of the solution (\ref{sandy}) with
$m=0$ plus something finite, whose precise value we will not need to know.
The point is that the boundary term $[\rho L']^\infty_0$ is {\it not} the same as the deficit angle.
From the asymptotic solutions at the origin and infinity we obtain:
$$-{4\pi^2\over e^2}[\rho L']^{r=\infty}_{r=0}\ =\ 0\ +\ 2\pi^2\vert mn\vert \lambda R^2T\ .$$
So the contribution of the vortex center to the action is indeed finite.
This differs from the situation for the simplified model studied in \cite{ichtop}, 
where the action was logarithmically divergent at the origin. 
Note that the flux tube action is also independent of $e$, which
means that we must still sum over Kaluza-Klein flux tubes in the 
semiclassical limit $e\rightarrow0$.
However, the vortices disappear (the vortex action becomes infinite) in the decompactification
limit $\kappa\rightarrow\infty$, i.e. $\kappa\rightarrow\infty$ \cite{witten11} of M-theory.

In the above we have imposed 6 constraints, but there are 7 free parameters
from four second-order equations with one first-order constraint. 
This leaves one modulus
of the vortex solution. This modulus is, roughly, the size of the vortex. 
To see this, note that without the cosmological constant (i.e. without the ``1'' on the RHS of
(\ref{one}-\ref{three})), the equations of
motion have the ``scaling'' symmetry
$$r\rightarrow pr\ \ ,\ \ \ \ L\rightarrow pL\ \ ,\ \ \ \ \rho\rightarrow p\rho\ \ ,
\ \ \ \ A\rightarrow{A\over p}\ \ ,\ \ \ \ f\rightarrow f\ \ $$
with an arbitrary real number $p$.
Of course, this symmetry is broken by the cosmological constant.

To summarize, the Kaluza-Klein flux tubes presented in this section for
the $3d$ gravity theory on the membrane have the following features: they
involve vortex lines of $x_{11}$,
they carry fractional magnetic charge
with respect to the Kaluza-Klein gauge field, 
they have finite action, and they are classically invisible outside their core.
These vortices are somewhat analogous to the Nielsen-Olesen vortices of the abelian
Higgs model \cite{nielsen} or to the Abrikosov flux tubes in a type II superconductor \cite{abrikosov},
with $x^{11}$ in the role of the Goldstone boson.

As stressed in the introduction, we should really consider a supersymmetric version
of the model presented here, which
has BPS-saturated (and therefore stable) vortices. This will be discussed elsewhere.

{}\subsection*{5. Sum over $3d$ topologies}\scs{5}

Let us now come to the geometric interpretation of fractional magnetic flux
(see \cite{ichtop} for some more details).
This flux has a geometric interpretation because the Maxwell field
is not just any gauge field, but the Kaluza-Klein gauge field
appearing in the line element
$$ds^2=dr^2+\rho^2d\phi^2+L^2(dz+\rho A d\phi)^2\ ,$$
where $(r,\phi,z)$ are cylinder coordinates.
Let us start with a two-brane of topology $S^2\times S^1$. Now assume that there is
magnetic flux
\ba\int_{S^2} F\ =\ 2\pi {m\over n}\ .\la{julchen}\ea
It is not difficult to see that this flux changes the topology of the manifold from $S^2\times S^1$
to the lens space $L(m,n)$.
As we know, we cannot define the gauge field such
that it is everywhere nonsingular. If we set it to zero at the South pole of the two-sphere,
then there is a Dirac string at the North pole, so the three-dimensional metric
will be singular there:
$$\rho A\ \rightarrow\ -{m\over n}\ \ \ \hbox{as}\ \ r\rightarrow 0\ .$$
This leads to some pathologies: e.g., the circle defined by
$r=\epsilon, z=0$ retains finite size $2\pi{mL\over n}$ as $\epsilon\rightarrow 0$.
On the other hand, the spiral line defined by
$$r\ =\ \epsilon\ ,\ \ \ - m\phi+nz\ =\ 0$$
shrinks to zero size as $\epsilon\rightarrow0$.
These pathologies can be removed by a large diffeomorphism.
By this we mean an $SL(2,Z)$ transformation
on the torus that is - for fixed $r$ - parametrized by the compact coordinates $z$ and $\phi$:
\ba
\phi&\rightarrow&\ \ \ a\phi+ bz\la{laura1}\\
z&\rightarrow&-m\phi+nz\la{john}\\
\tau &\rightarrow&\ {\ \ n\tau+m\over -b\tau+a}\ ,\la{laura3}\ea
where
$$\tau\ = \rho A\ +\ i{\rho\over L}$$
is the modular parameter of the torus and
$a,b$ are integers such that $an+bm=1$. 
The latter condition ensures that the determinant of the map is one,
while the condition $a,b,m,n\in Z$ ensures that closed lines are mapped onto closed lines.

This $SL(2,Z)$ transformation leaves the line element invariant
and removes the singularity of the Klauza-Klein gauge field.
But it changes the topology of the membrane. More precisely,
a manifold on which a metric that corresponds to fractional magnetic
flux (\ref{julchen}) can live is constructed as follows. One decomposes the
two-sphere into two coordinate patches $D_1$ and $D_2$, where we can choose $D_1$
to be a little disc around the North pole and $D_2$ to be its complement in $S^2$.
One cuts out the solid torus $D_1\times S^1$ from the three-manifold $S^2\times S^1$,
twists it by the $SL(2,Z)$ transformation and glues it back in, such that its
meridian $M'$ gets identified with a spiral line on the surface of the $D_2\times S^1$:
$$M'\ =\ nM+mL\ .$$
Here, $M$ is the meridian and $L$ is the longitude of the torus that bounds $D_2\times S^1$.
This operation of cutting a tube out of a three-manifold,
twisting it as described and gluing it back in
is called Dehn surgery.
$m$ and $n$ are called Seifert invariants; they must be relatively prime.
It can be shown that the values of $a,b$ do not influence the topology of the
obtained manifold. The three-manifolds constructed
in this way are by definition the lens spaces $L(m,n)$.
They definitely differ from $S^2\times S^1$: e.g., their first homology group is finite, $Z_m$.

As a first example, consider the case $n=1$. Then we can choose $a=1,b=0$, so
the large diffeomorphism (\ref{laura1}-\ref{laura3}) is:
$$\phi\rightarrow\phi\ ,\ \ \ \ z\rightarrow z-m\phi\ ,\ \ \ \ \rho A\rightarrow\rho A+m\ .$$
Those are the large gauge transformations of an ordinary $U(1)$ gauge theory, if
the membrane is interpreted as the total space of the $U(1)$ bundle over 
$S^2$ that is represented by the compact coordinate $z$. $m$ is the obstruction
to the existence of a section of the $U(1)$ bundle, and
$M'$ is identified with $M+mL$. 

As an example of a large gauge transformation
in Kaluza-Klein gauge theory that does not have an analog in ordinary gauge theory,
consider the case $n=5,m=-2$. Then one can pick $a=1,b=2$.
In the old coordinate system, $L=5$ and $\rho=r$. Then the modular parameter transforms as
$$\tau\ =\ {2\over5}+{i\over5}r\ \rightarrow\ {-5ir}\ +\ O(r^2)\ =\ \tau'\ ,$$
which corresponds to $L=1, \rho=5r$ in the new coordinate system (flipping an orientation).
The resulting manifold is the Lens space $L(5,2)$; the deficit (or rather surplus) angle
$-8\pi$ does not change the topology of the manifold further. 

As in \cite{ichtop}, by performing
Dehn surgery with arbitrary Seifert invariants along an arbitrary number of
vortex lines running around the $S^1$, and by also replacing the $S^2$ by
an orientable surface of arbitrary genus, or by an unorientable surface
(in this case the $S^1$ must be replaced by a circle bundle that reverses
orientation along closed lines on the surface along which the surface reverses
orientation), one can construct all possible orientable three-dimensional topologies
that admit a foliation by circles (Seifert manifolds \cite{seifert}).
Those also include the three-sphere (which admits infinitely many ``Hopf fibrations'').
So the sum over Kaluza-Klein flux tubes implies a sum over membrane topologies! 

For completeness, let us mention how one can construct {\it all} three-dimensional
topologies: one starts with an $S^3$ and draws an arbitrary knot, or a set of linked knots
on it. One performs Dehn surgery along the knot lines by cutting out tubes around
them, then twists each tube with arbitrary
Seifert invariants $m,n$ as described above and glues it
back in. The statement is that if one draws all possible knots and links and performs all
possible surgeries on them, then one obtains all possible topologies.
This does of course not classify three-dimensional topologies, because, first,
one still has to classify knots, which is an unsolved problem; and second,
even if one starts with different knots and performs different surgeries, one
might still end up with the same topology. But in one way or another, all possible
topologies occur.

{}\subsection*{6. Confined Kaluza-Klein momentum}\scs{6}

Let us now consider
the Wilson loop for the Kaluza-Klein gauge field on a membrane of initial 
topology $R^2\times S^1$, keeping the two-dimensional meric on the $R^2$
fixed:
$$ \exp\{W(q,C)\}\ =\ <\exp\{iq\oint_CA_\alpha d\sigma^\alpha\}>\ .$$
Here $C$ is a closed contour in $R^2$ and $q$ is a test charge.
We assume that the semiclassical 
limit $e\rightarrow0$ has been taken; so what we mean by the 
brackets on the right-hand side is an average over classical solutions,
including an 
arbitrary numbers of Kaluza-Klein flux tubes (or ``vortices'') running through the loop.
In the absence of flux tubes, the gauge theory is Higgsed and the Wilson loop
obeys a perimeter law. In order to argue that flux tubes turn this into an
area law, we adapt a well-known argument from the two-dimensional Abelian Higgs model
(see, e.g., \cite{callan}) to the case at hand.

As seen in section 3, the flux tubes are classically invisible outside their core,
and in particular there is no deficit angle. So let us assume that we can approximate the 
system of flux tubes by a dilute gas of noninteracting loops that are wrapped 
around the $S^1$ and carry magnetic flux $2\pi{m\over n}$.
Flux tubes that run through the inside of the loop $C$ contribute an 
Aharonov-Bohm phase exp($\pm 2\pi iq{m\over n}$) to the Wilson loop.\footnote{We
mean flux tubes that are inside the Wilson loop and are linked with it by running around the $S^1$.
Flux tubes that are not linked with the loop do not contribute a phase.}
In the path integral, each vortex also comes with a weight factor 
$e^{-S_{n,m}}$, where $S_{n,m}$ is the finite action of a single $(n,m)$ vortex.
A vortex also comes with a factor 
$${\hbox{Area inside $C$}\over a^2} \ ,$$
which is the number of possible vortex locations inside the loop\footnote{We should
also sum over the possible shapes of vortex lines. We neglect this;
this is like describing the flux tubes in
a $3d$ superconductor by the Nielsen-Olesen vortices in a $2d$ abelian Higgs model (which works).}
($a$ is a length scale such as
a short-distance cutoff on $R^2$).
Consider first the partition function $Z_{n,|m|}$ of the gas 
of $(n,\pm |m|)$ vortices inside $C$ in the presence of the Wilson loop,
with $n,|m|$ held fixed:
\ba Z_{n,|m|}(q,C)&\sim& 
\sum_{N_+,N_-=0}^\infty
 {1\over N_+!N_-!} ( e^{-S_{n,m}}\ {\hbox{Area}\over a^2})^{N_++N_-}
\ e^{ 2\pi iq{m\over n}(N_+-N_-)}\\
&=& \sum_{N=0}^\infty{1\over N!}
\{ e^{-S_{n,m}}\ {\hbox{Area}\over a^2}\ (e^{ 2\pi iq{m\over n}}+e^{-2\pi iq{m\over n}})\}^N\\
&=& \exp\{2\cos(2\pi{m\over n}q)\ {\hbox{Area}\over a^2}\ e^{-S_{n,m}}\}\ .\ea
Here $N_+$ and $N_-$ are, respectively, the number of $(n,|m|)$ 
and $(n,-|m|)$ vortices inside the loop, and $N=N_++N_-$.
This yields, up to perimeter terms, the Wilson loop
\ba 
W_{n,m}(q,C)\ \ =\ \ \log {Z_{n,|m|}(q)\over Z_{n,|m|}(q=0)}\ \ =\ \ 
\sigma_{n,m}\ \times\ \hbox{Area inside $C$}\ ,\la{square} 
\ea
where the string tension 
\ba\sigma_{m,n}\ \sim\ {2e^{-S}\over a^2}\ (\cos\ 2\pi{m\over n}{q}\ -\ 1)\ \la{triangle}\ea
is periodic in $q$ with period $n$. $W(q,C)$ is simply the sum over the $W_{n,m}$.
The string tension is proportional to the
vortex density, which diverges as ${1\over a^2}$ since our vortices have finite action
and are therefore always in a condensed phase.\footnote{This differs from 
the situation in the toy model studied in 
\cite{ichtop}, where the vortex action was logarithmically divergent; this led to an
anomalous dimension of the vortex density and to a phase transition.}
We see that the condensation of flux tubes with given $n$ leads to linear confinement of
electric charges $q$, unless $q$ is an integer multiple of $n$.
Electric charges that are a multiple of $n$ are screened rather than linearly confined
(the string tension is zero). 

But what is electric charge in our context? The electric charge that couples to the 
Kaluza-Klein gauge field is nothing but Kaluza-Klein momentum: 
the Kaluza-Klein modes 
\ba x_{q}(\sigma_1,\sigma_2)e^{iqz}\la{star}\ea
appear in the two-dimensional gauge theory
as scalar fields with charge $Q=q$ and mass $M=q$, since in (\ref{diana}):
$$|\partial_ix_{q}|^2\ \ \rightarrow\ \ 
|(\partial_\alpha \ -\ iqA_\alpha) x_{q}|^2\ +\ {q^2\over L^2}\ x_{q}^2\ .$$
So the linearly confined charge is Kaluza-Klein momentum in the circular 
membrane direction.

Before we speculate about what this means, let us compare with previous 
suggestions
in the context of Kaluza-Klein compactification from four to three dimensions. 
In this situation one has a $3d$ $U(1)$ Kaluza-Klein gauge field, and there exist Kaluza-Klein 
instantons (instead of Kaluza-Klein flux tubes).
Those are the Kaluza-Klein monopoles of \cite{sorkin}, dimensionally reduced.

In ordinary $3d$ compact $U(1)$ gauge theory, a Coulomb plasma of instantons and
anti-instantons forms that leads to
linear confinement of electric charge \cite{polyakov3d}. In \cite{gross}, it was asked whether 
Kaluza-Klein momentum might also be linearly confined due to the
condensation of Kaluza-Klein instantons, and a problem was pointed out:
In an ordinary $3d$ gauge theory, there is an attractive Coulomb potential between instantons
of opposite magnetic charges, and a repulsive potential between instantons of like charges. 
But while in a Kaluza-Klein gauge theory there is still an attractive Coulomb potential
between instantons of opposite magnetic charges, it can be seen that
there is no potential between instantons of like charges. Can a Coulomb plasma still form?

A comparison with
$N=2$ supersymmetric Yang-Mills theory \cite{seibergwitten} suggests that the answer is ``No":
At the $N=2$ supersymmetric point the magnetic monopoles are BPS and
there is an attraction between opposite magnetic charges,
but no repulsion between like magnetic charges. 
The fact that there is no confinement at the $N=2$ point confirms that
no stable monopole-antimonopole plasma can form.
When one perturbs to $N=1$, one also creates a small repulsion between
monopoles of like magnetic charges. Then a stable monopole-antimonopole plasma forms
that linearly confines electric charge.\footnote{I thank A.M. Polyakov for pointing out
this interpretation.}

Thus, confinement of Kaluza-Klein momentum due to the condensation
of Kaluza-Klein instantons does not quite seem to work.
But in compactification from 3 to 2 dimensions, there is no similar problem!
There is neither an attractive nor a repulsive force between the Kaluza-Klein flux tubes.
So as vortices in a $2d$ abelian Higgs model, vortices in the Kaluza-Klein model can still
form a dilute gas that leads to linear confinement.

{}\subsection*{7. Interpretation}\scs{7}

Let us now offer a speculative interpretation of the result (\ref{square}-\ref{triangle}).
The total string tension is simply 
the sum $\sum_{m,|n|}\sigma_{m,|n|}$ of the individual string tensions.
So it seems that {\it all} Kaluza-Klein momenta are linearly confined.
As usual, this should imply that all states in the physical Hilbert space are neutral under
global symmetry transformations. But here, those are just translations
in $z$-direction. Thus, at large scales, the ``matter fields'' should depend 
only on $\sigma_1$ and $\sigma_2$, and not on $z$.
We conclude that the membrane is effectively two-dimensional in embedding space. 
One could perhaps say that the membrane is ``permanently Kaluza-Klein compactified'', similarly as
quarks are permanently confined in QCD.

This does not yet imply that the Kaluza-Klein modes (\ref{star}) are completely gone.
In ordinary gauge theory, there can be neutral bound states of charged fields.
Can there be bound states of Kaluza-Klein modes with zero total Kaluza-Klein momentum 
(``baryons'')?

In ordinary gauge theory, neutral bound states can exist because opposite electric charges attract
and like charges repel, so there are no net long-range forces between neutral
pairs.
But while in Kaluza-Klein gauge theory opposite electric charges still attract,
there is no force between particles of equal charge.
This can easily be seen by computing the classical forces between
strings of matter that are wrapped around $z$ and rotate in $z$-direction at the speed of light
(those are the three-dimensional objects that reduce, upon Kaluza-Klein compactification,
to electrically charged particles).

This might mean that in Kaluza-Klein gauge theory free baryons cannot exist,
because the attractive and repulsive forces between their constituents no longer balance each other.
So after integrating out $A_\alpha$ and $x_{11}$,
the Kaluza-Klein modes might disappear completely and what remains might be
standard strings with embedding coordinates $x^\mu(\sigma_1,\sigma_2)$.
This might be exactly what is needed to
make sense out of M-theory, as the statistical mechanics of fluctuating membranes
that have a continuum limit in which they are ``wrapped'' 
down to strings.
Of course the strings would be very foamy from a $3d$ viewpoint.

In 3+1 dimensions, the
sum over spacetime topologies has been made responsible for various phenomena,
such as the loss of quantum coherence or the vanishing of the cosmological constant.
Our sum over $3d$ topologies that admit a foliation by circles suggests neither of these effects,
but instead ``confinement of Kaluza-Klein momentum'' as an equally dramatic one: perhaps
some sense can be made out of dynamical gravity in more than two dimensions after all -
namely, that it dynamically reduces to renormalizable $2d$ gravity.

The discussion can be generalized to $p-$branes with $p>2$.
The Dirichlet $p$-brane contains a world-brane gauge field \cite{polchinski},
whose dual is a $(p-2)-$form.
The field strength of this $(p-2)$-form can be integrated over a $(p-1)-$ cycle $K$ on the world-brane.
There could be Kaluza-Klein flux tubes of the nonabelian Kaluza-Klein gauge theory that
is obtained by compactifying the $p-$brane on $K$. Summing over them would correspond to
summing over $(p+1)-$dimensional topologies that can be foliated by $K$. Criteria for
confinement might translate into statements about when branes wrap and when they don't.

\subsection*{Acknowledgements}
I thank the members of the Theory groups in Santa Barbara, Pasadena, Princeton, Bern,
Z\"urich, Potsdam and Berlin for questions, discussions and hospitality.
In particular, I thank J. Fr\"ohlich, D. Gross, M. Leibundgut, D. L\"ust,
H. Nicolai, M. Niedermaier, A. Polyakov and J. Schwarz for conversations.

{}

\end{document}